# Radiation Dose Risk and Diagnostic Benefit in Imaging Investigations

## Lidia Dobrescu[1], Gheorghe-Cristian Rădulescu[2]

[1]Electronics Telecommunications and Information Technology Faculty, University Politehnica of Bucharest, Bucharest, Romania
[2]Radiology Department, Central Military Universitary Emergency Hospital, Bucharest, Romania

**Email address:**
lidia.dobrescu@electronica.pub.ro (L. Dobrescu), radulescutcristi@yahoo.com (Gheorghe-Cristian Rădulescu)

**Abstract:** The paper presents many facets of medical imaging investigations radiological risks. The total volume of prescribed medical investigations proves a serious lack in monitoring and tracking of the cumulative radiation doses in many health services. Modern radiological investigations equipment is continuously reducing the total dose of radiation due to improved technologies, so a decrease in per caput dose can be noticed, but the increasing number of investigations has determined a net increase of the annual collective dose. High doses of radiation are cumulated from Computed Tomography investigations. An integrated system for radiation safety of the patients investigated by radiological imaging methods, based on smart cards and Public Key Infrastructure allow radiation absorbed dose data storage.

## 1. Introduction

People are exposed to natural radiation from ground, water, air or cosmic radiation, called background radiation. The general level of this type of radiation is still usually low.

In our modern life a great concern about radiation is present due to modern irradiative sources such as mobile phones or great antennas. Nuclear accidents anywhere in the world and the general perception on cumulative radiation doses increase the gravity of the problem.

Medical investigations are major contributing factors, but still not so well perceived.

The absorbed radiation dose and their implications for human health are still highly controversial.

The International Commission on Radiological Protection (ICRP) recommends that the public limit of artificial irradiation should not exceed an average of 1 mSv effective dose per year, not including medical and occupational exposures.

ICRP limits for occupational workers are 20 mSv per year, averaged over defined periods of five years, with the further provision that the dose should not exceed 50 mSv in any single year [1].

The radiographies, CT-s and generally X-rays investigations can save life but their high level radiation doses can affect people health. More and more patients are investigated by radiographies and CTs and these kind of radiological methods strongly increase the cumulative radiation dose received by patients. Chest/Torax, Cervical, Thoracic and Lumbar Spine Radiographies and CT for head, neck, chest, spine and abdomen are common investigations performed in many countries.

In order to record different types of investigations and their individual radiation absorbed dose, worldwide current medical practice uses paper forms, files and folders.

Modern radiological apparatus for computerized tomographies or scintigraphies can provide the radiation doses during a particular investigation, but the recorded doses' types and the radiation measurement units in different types of investigations are not the same.

The reporting system is not accurate because the individual doses recorded by patients are not cumulated. The patients travel from one hospital to another, all over the country or all over the world.

The total volume of prescribed medical investigations, starting from minor dental radiographies and ending at major CT scans prove a serious lack in monitoring and tracking of the cumulative radiation doses in many health services all over the country and their received radiation doses are quit impossible to be cumulated.

Modern radiological investigations equipment is continuously reducing the total dose of radiation due to improved technologies, so a decrease in per caput dose can be noticed, but the increasing number of investigations has determined a net increase of the annual collective dose.

So, dose reduction technology provided by new radiological equipment is not enough. New approaches of managing the dose for patients and for healthcare systems are widely debated.

In spite of these general doses reduction concern, overlapped exposures are still often used for a sure and precise diagnosis.

A reasonable balance between the applied medical radiation and the image quality must be always kept, towards a proper radiological risk versus the diagnostic gain.

Early and detailed diagnoses for prevention and therapy of suspected diseases imply repeated and overlapped exposures.

A secure integrated system is designed in a new project. The new system is designed on smart cards technology. Integration of PKI infrastructures supplies a high level of security for the whole system including access to databases through various applications and it also ensures the confidentiality of citizens' personal data stored on cards and in a central data base.

## 2. Radiation Doses and Conversion

The International System of Units (SI) uses for measuring the equivalent absorbed radiation dose, the Sievert (Sv) as a derived unit. This is the central unit of the implemented project.

There are many other different units for absorbed radiation commonly used.

The SI radiation units can be used for three different purposes.

The first one is the radioactivity released by a material source. There are many radioactive materials that can emanate radiations. For all of them the appropriate units are the Becquerel (Bq) and the Curie (Ci) as SI units.

The second class includes the exposure, monitoring the total amount of radiation travelling generally through the air. The Coulomb/Kilogram (C/kg) and the Roentgen (R) are commonly used measurement units. Many radiation monitors display the dose using these two units.

The third class refers to the absorbed dose. The Gray (Gy) is widely used and the Rad is another measurement unit used with quantities of absorbed dose.

There is one more class, that was not considered as an independent one, that mix together the amount of radiation absorbed and the biological effects of that type of radiation and the specific measurement units are Roentgen equivalent man (rem) and Sievert (Sv).

The Gray from the third class can describe any material, while the Sievert better describes the effective and committed equivalent dose absorbed by biological tissues.

The commonly used doses for medical investigations are from different classes. For scintigraphies common doses are expressed in MBecquerels, for Computed Tomographies the usual doses are expresssed in mGy*cm2 and for normal radiographies the dose is commonly expressed in mGy. So an integrated system must convert and unify them.

A particular situation was determined by CT-s recorded radiation doses. The radiation dose provided by modern electronic equipments is expressed in two related measurement systems: CTDI (CT dose index) and in DLP (dose length product).

DLP was chosen as the major input data for the system in CTs investigations but the reported data must reach the biological effective dose. Conversion factors can be used. However, these conversion factors are problematic in that they are only estimators of doses and do not represent the full range of pediatric sizes [2]. The conversion factors can slightly vary from different manufacturers [3].

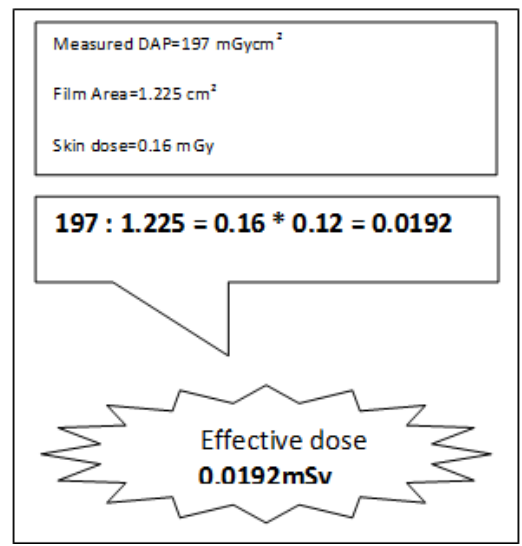

*Figure 1. Effective dose calculus in a radiological investigation.*

Typically CTDI doses are 15-20 mGy, higher than the usually absorbed doses from common radiographies.

For classic radiological investigation, an important topic for data management is the measurement and the calculation

In diagnostic X-ray examinations dose area product (DAP) is linked with radiation risk measurements. It can be calculated by multypling the absorbed dose and the irradiated area. Its Measurement units is (Gy*cm2). DAP reflects not only the dose within the radiation field but also the area of tissue irradiated.

It also has the advantage of being easily measured, with a DAP meter on the X-ray set. DAP meters are available apparatus in radiology units measuring absorbed doses. DAP meters are linked with the absorbed dose to air.

DAP reflects the dose and the irradiated tissue area. It can indicate the overall risk of inducing cancer. The irradiated area increases with the square of distance from the source due to the beam divergence. The radiation intensity decreases with the inverse square of distance. So their product becomes independent of distance.

More refinements can be used in order to consider attributable lifetime risk. The multiplicative risk projection model can be different for males and females, always relatively higher for females. Over 80 years of age, the risk can be neglected. The younger people are more radiosensitive.

A DAP calculus example is shown in figure 1 where the biological tissue factor for a lung radiography was considered to be 0.12 [4].

Special conversion factors are also available for children of various ages, but this project does not include children.

The Sievert (Sv) as an International System of Units (SI) derived unit of radiation dose was chosen as the central unit of the project.

## 3. Radiation Effects

The critical elements in human cells are DNA molecules involved in the process of repairing the damages. They may be successful in their repairing mission or they may be sacrificed. Sometimes, altered DNA can result and the mutants will also disappear. But there is a small possibility of mutants to survive and his can be the start of a multi-step process that could eventually lead to formation of a cancer [5].

Chromosomal aberrations or aneuploidy from persistent DNA damage may lead to genomic instability.

The low doses of ionizing radiation biological effects are controversial. Radiation risk evaluation methods and models were developed.

The International Commission on Radiological Protection (ICRP) recommends the use of the LNT model [6].

The most popular linear no-threshold LNT model assumes proportionality between dose and cancer risk with a linear dependence from 1 mGy to 100 Gy.

This model can be suitable for small-dose exposures and it is still divisive. The threshold model assumes that very small exposures are harmless.

Another radiation effect, Hormesis, explains that very small doses of radiation can be beneficial. The adaptive responses of the human cells were observed at low doses and disappear with higher doses [7]. This protective mechanism at the cell and tissue levels has been described in many experiments. It operates against cancer developing mechanisms. A dose of 10 mGy reduces the rate of spontaneous transformation in culture cells below the background level. Epidemiologic studies suggest that Hormesis also exists in human cells.

Radiation effects can be deterministic or stochastic, prompt or delayed, somatic or genetic.

An important distinction between stochastic and deterministic effects has to be analyzed. It can be accepted that the future incidence of cancer can increase with the radiation dose and there are many quantitative models that try to predict the level of risk linked with the absorbed dose.

A stochastic effects means that its occurrence probability increases with the total absorbed dose, while its severity is stochastic effect. No threshold dose can be established for stochastic effects. Induced cancer can be regarded as a stochastic effect.

The other types of effects are the deterministic ones, as shown in figure 1. Usually over the threshold of 10 Sv, death and severe health effects are always present. The risk of radiation exposure has a huge importance in the case of an emergency situation such as a nuclear accident, when high doses of radiation are absorbed leading to deterministic effects. The consequences of these situations are harmful effects in a direct ratio to the exposure.

Commonly in fact it is hard to distinguish between the deterministic and stochastic effects.

Compared to classical radiography, CT is a high-dose imaging method, although doses are still below the threshold dose for deterministic effects.

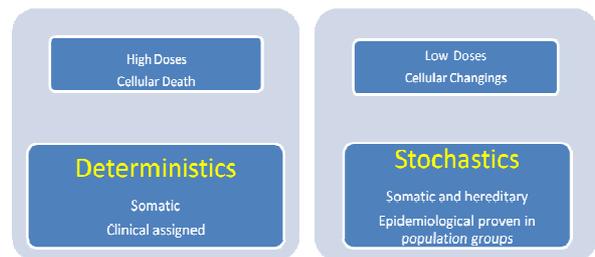

*Figure 2. Radiation biological effects.*

## 4. Experimental Results

Cumulating the total effective dose for each patient provides a more accurate method to generate reports.

The new system, designed on smart cards technology covers one major need of the health-care system. Such a radiation safety system can provide a couple of secure services like electronic record of patient's radiological investigations, assistance in prescription of future radiological investigations based on patient's history and different reports and statistics. The smart cards allow authentication, digital signature and secure data storage.

The system provides the replication of the information stored in central databases, local databases and patient cards. An initial performed pilot survey has revealed many cases of over passing the maximum cumulative dose only during one single hospitalization. The data has been collected for computerized tomography and scintigraphy and no by classical radiological methods. For a computed tomography the radiation dose can be ten times greater than in chest radiography. Chest X ray is considered as the standard reference in comparing radiation doses and it brings 0.02 mSv. The usage of two-dimensional images of the distribution of radioactivity in tissues after the internal administration of a radiopharmaceutical imaging agent, increases the cumulative dose.

From 1177 patients from one single unit that have been investigated using radiological procedures such as radiography, CT scans and scintigraphies, 1162 patients were radiologically investigated, 78 were scanned by computed tomographies and 15 patients were analyzed by scintigraphies.

Only 4 of them had high doses investigations consisting in mixed CT scans and scintigraphies and 3 of them overpass the maximum allowed cumulative dose that was initially set at 20 mSv, but 35 of them overpass the maximum allowed dose only from CT scans.

The maximum effective dose in a performed scintigraphy was 5.920 mSv.

The typical effective doses from CT scans are exposed in figure 3.

Their measurement unit is the Sievert.

The values were extracted from a study for 36 European countries, including Romania [8].

Our determined medium values for CT investigations were sometimes higher.

| Investigation type | Effective Dose (mSv) |
|---|---|
| Chest/Thorax | 0.1 |
| Cervical Spine | 0.2 |
| Thoracic Spine | 0.6 |
| Lumbar Spine | 1.2 |
| Mammography | 0.3 |
| Abdomen | 0.9 |
| Pelvis and Hip | 0.7 |
| Ba meal | 6.2 |
| Ba enema | 8.5 |
| Ba follow-through | 7.2 |
| IVU | 2.9 |
| Cardiac angio-graphy | 7.7 |
| CT head | 1.9 |
| CT neck | 2.5 |
| CT chest | 6.6 |
| CT spine | 7.7 |
| CT abdomen | 11.3 |
| CT pelvis | 7.3 |
| CT trunk | 14.8 |
| PTCA | 15.2 |

*Figure 3. Average values of typical effective doses for TOP 20 groups.*

There are still uncertainties in the accuracy of the coefficients used to convert the measured dose quantities into typical effective doses.

## 5. Managing Radiation Doses for Diagnostic Benefit

Generally the radiological imaging procedures are carried out at the request of the treating physician and the radiologist.

Their prescription is based on several objectives and subjective elements:
- Therapist request
- Patient history
- Imaging investigation algorithm
- The results of the investigations already carried out
- The personal experience of the examiner

In the practice of medicine, there must be a judgment made concerning the benefit/risk ratio. This requires not only knowledge of medicine but also of the radiation risks.

The aim of managing radiation exposure is to minimize the irradiative risk without sacrificing, or unduly limiting, the obvious benefits in the prevention, diagnosis and also in effective cure of diseases (optimization). It should be pointed out that when too little radiation is used for diagnosis or therapy there is an increase in risk although these risks are not due to adverse radiation effects per se. Too low an amount of radiation in diagnosis will result in either an image that does not have enough information to make a diagnosis.

There are several ways that will minimize the risk without sacrificing the valuable information that can be obtained for patients' benefit. Among the possible measures it is necessary to justify the examination before referring a patient to the radiologist or nuclear medicine physician.

Repetition should be avoided of investigations made recently at another clinic or hospital.

An investigation may be seen as a useful one if its outcome - positive or negative - influences management of the patient. Another factor, which potentially adds to usefulness of the investigation, is strengthening confidence in the diagnosis.

Most common examples of unjustified examinations include: routine chest radiography at admission to a hospital or before surgery in absence of symptoms indicating cardiac or pulmonary involvement (or insufficiency); skull radiography in asymptomatic subjects of accidents; lower sacrolumbal radiography in stable degenerative condition of the spine in the 5th or later decade of life, but there are of course many others [9].

From medical experience in radiological hospital unit the diagnosis benefit is a priority.

For a lung Rx examination any suspected image is repeated or supplementary investigated in a Computed Tomography.

Many times, the personal experience of the radiologist is the dominant factor in high radiation doses investigations:
- Urgent exam such as trauma and polytrauma, fractured skull or spine where CT exams are a real benefit for the patient and an economic benefit for the hospital. The shortening of the patients hospital stay
- scintigraphic investigation in thyroid diseases
- repeated mammography
- bone disease with seemingly benign tumors requiring an examination for differential diagnosis

## 6. Conclusions

The initial performed pilot survey has revealed many cases of over passing the maximum cumulative dose that was initially set at 20mSv only during one single hospitalization.

High doses of radiation were cumulated from Computed Tomography investigations.

The abdomen and pelvis CT scans have often provided effective doses over the maximum allowed dose[10].

Conversion problems were detected for multiple CT scans in one single examination, because more DLP are measured in the same examination and recording them with different conversion coefficients becomes a long procedure.

Computed tomography alone is responsible for high doses of radiations.

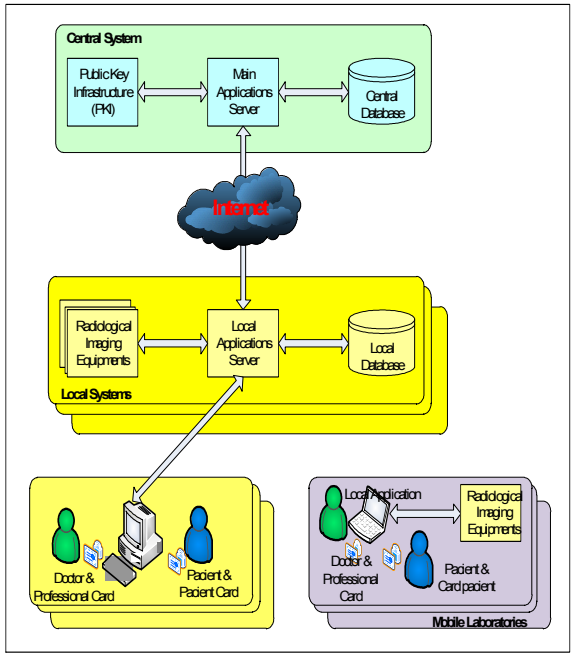

*Figure 4. The architecture of the integrated system*

- Many times the diagnosis accuracy imposes overlapped medical procedures. There are radiological procedures with high effective doses
- A special position is occupied by computed tomography (CT), and particularly its most advanced variants like spiral or multi slice CT. Usefulness and efficacy of this great technical achievement is beyond doubt in particular clinical situations, however the ease of obtaining results by this mode and temptation to monitor frequently the course of a disease or perform screening should be tempered by the fact that repeated examinations may deliver an effective dose of the order of 100 mSv, a dose for which there is direct epidemiological evidence of carcinogenicity.
- A pilot integrated system for radiation safety and security of the patients investigated by radiological imaging methods such as radiographies, computed tomographies or scintigraphies is realized. The system is based on smart cards and Public Key Infrastructure. The implementation of system's architecture has three distinct levels of storage: a central database, many local databases and Citizen Radiation Safety Cards. The smart cards allow authentication, digital signature and secure data storage.
- The system provides the replication of the information stored in central databases, local databases and patient cards to cover any unusual possible situation such as:

- The patient goes to the doctor without the patient card.
- The patient goes to the doctor with no card and the hospital unit's information system does not have access to the central database (ex: for mobile laboratories)
- The patient presents the card to the doctor but the doctor does not have access to any database (local or central).

## Acknowledgements

The SRSPIRIM 187/2012 project in Romanian Collaborative Applied Research Projects Subprogram has supported the research work for this article. The authors wish to address thanks to all the persons involved in this project for their support, work and ideas.